\documentclass[twocolumn,aps,prl,groupedaddress]{revtex4}
\usepackage{amssymb}
\usepackage{amsmath}
\usepackage{color}
\usepackage{graphicx}
\usepackage{marvosym}
\usepackage[normalem]{ulem}

\newcommand{\be}{\begin{equation}}
\newcommand{\ee}{\end{equation}}
\newcommand{\nn}{\nonumber\\}

\newcommand{\p}{\partial}
\newcommand{\la}{\langle}
\newcommand{\ra}{\rangle}
\newcommand{\lb}{\left[}
\newcommand{\rb}{\right]}
\newcommand{\lp}{\left(}
\newcommand{\rp}{\right)}

\renewcommand{\Re}{{\rm Re}\,}
\renewcommand{\vec}[1]{{\bf #1}}

\begin{document}
\title{Symmetry, spin-texture, and tunable quantum geometry in a WTe$_2$ monolayer}
\author{Li-kun Shi$^{1}$ and Justin C. W. Song$^{1,2}$}
\affiliation{$^1$Institute of High Performance Computing, Agency for Science, Technology, \& Research, Singapore 138632}
\affiliation{$^2$Division of Physics and Applied Physics, Nanyang Technological University, Singapore 637371}

\begin{abstract}
The spin orientation 
of electronic wavefunctions in crystals is an {\it internal} degree of freedom, typically insensitive to electrical knobs. 
We argue from a general symmetry analysis and a $\vec k \cdot \vec p$ perspective, that monolayer 1T'-WTe$_2$ possesses a gate-activated canted spin texture that produces an electrically tunable bulk band quantum geometry.
In particular, we find that due to its out-of-plane asymmetry, an applied out-of-plane electric field breaks inversion symmetry to induce both in-plane and out-of-plane  
electric dipoles. These in-turn generate spin-orbit coupling to lift the spin degeneracy and enable
a bulk band Berry curvature and magnetic moment distribution to develop. 
Further, due to its low symmetry, Berry curvature and magnetic moment in 1T'-WTe$_2$ possess a dipolar distribution in momentum space, and can lead to unconventional effects such as a current induced magnetization and quantum non-linear anomalous Hall effect. 
These render 1T'-WTe$_2$  a rich two-dimensional platform for all-electrical control over quantum geometric effects.
\end{abstract}

\pacs{pacs}
\maketitle

Structure and material property/functionality have an intimate relationship. A striking example is monolayer WTe$_2$ where a structural change from 1T to a distorted 1T' structure induces a topological phase transition from trivial to $Z_2$ topological phase~\cite{Qian}. Recently realized in experiment \cite{Tang,Fei,Wu}, the distorted 1T'-WTe$_2$ monolayer possesses a large bulk bandgap $\sim 0.055\, {\rm eV}$~\cite{Tang}, and helical edge modes that mediate robust edge conduction~\cite{Fei,Wu} characteristic of a robust quantum spin Hall state. 

\begin{figure}
\includegraphics[scale=0.21]{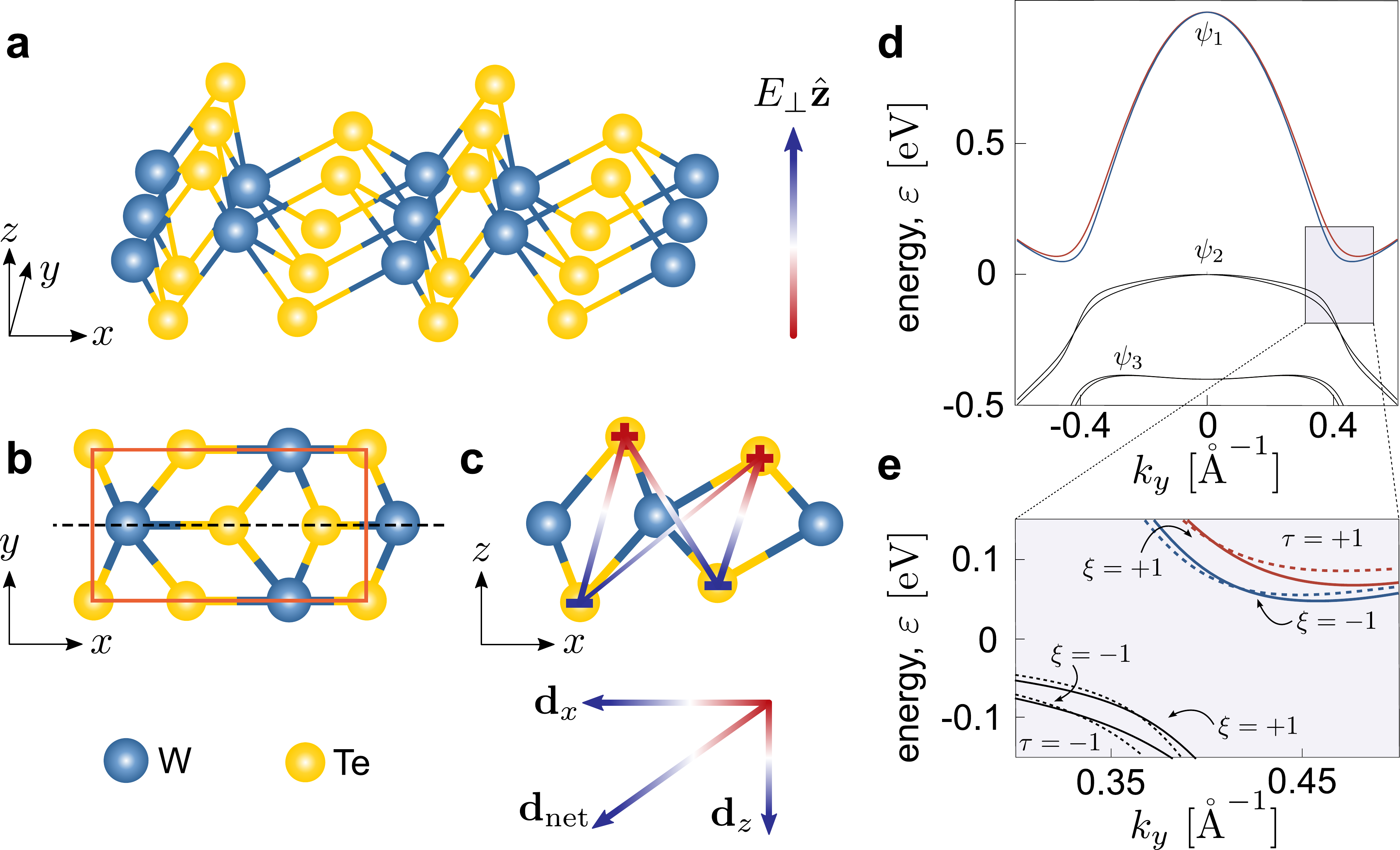}
\caption{ 
({\bf a,b}) Crystal structure for a 1T'-WTe$_2$ monolayer possess a particularly low symmetry with ({\bf b}) a single mirror plane black dashed line (primitive cell denoted by red box). 
({\bf c}) A net in-plane dipole moment $d_x$ can be induced by a perpendicular electric field $E_\perp$ as a result of non-alignment of the Te atoms on the top and bottom layers. ({\bf d,e}) The electronic band structure of bulk 1T'-WTe$_2$ monolayer along $k_y$ direction when an perpendicular electric field is applied possesses 
(zoom-in, {\bf e}) spin split conduction and valence bands near the gap opening. Solid and dashed lines are bandstructure from six-band (SBD) and effective four-band $h^{\rm eff} (\vec k)$ models respectively. 
Parameters used: for the pristine part we used values listed in Table~\ref{tab:parameters} \cite{Supp}. For the electric field induced part we used $\alpha_{x,y} = \lambda = \delta_x = 0$, and $\delta_z = 0.025~{\rm eV}$ as an illustration.
}
\label{fig1}
\end{figure}

\begin{figure}
\includegraphics[scale=0.22]{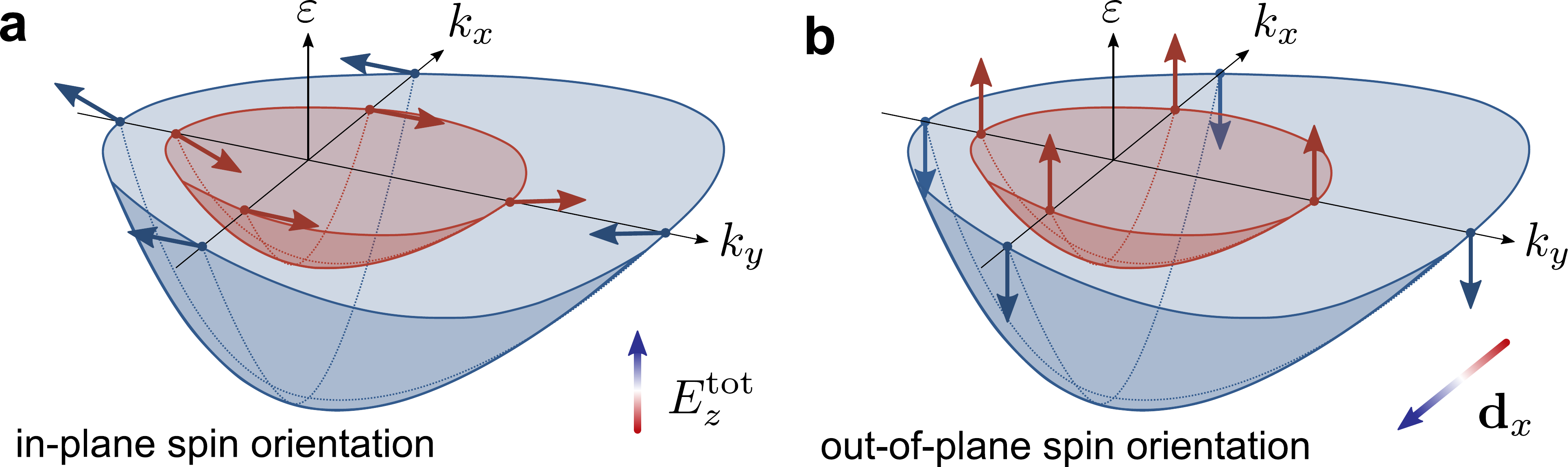}
\caption{ ({\bf a},{\bf b}) Schematic spin splitting and spin texture of the conduction band (near the gap)
induced by an applied electric field. These can be classed as 
({\bf a}) 
In-plane spin orientation
($\alpha, \delta_x \neq 0$, $\lambda,\delta_z = 0$) originating from an out-of-plane electric field $E_z^{\rm tot}$, or
({\bf b}) 
Out-of-plane spin orientation
($\alpha,\delta_x = 0$, $\lambda, \delta_z \neq 0$) arising from an in-plane dipole $d_x$. 
}
\label{fig2}
\end{figure}

Here we argue that, aside from determining the band topology, the distorted crystal structure of 1T'-WTe$_2$ (Fig.~\ref{fig1}a-c) also enables unusual bulk band quantum geometry and spin physics to be accessed and controlled.
By developing a low energy $\vec k \cdot \vec p$ model from symmetry analysis we find that when an out-of-plane electric field $E_\perp$ is applied, spin-degeneracy is lifted (Fig.~\ref{fig1}d,e) by inducing both in-plane 
as well as out-of-plane 
 spin orientations (Fig.~\ref{fig2}a,b).  
While in-plane 
spin orientations are synonymous with an out-of-plane inversion symmetry (IS) breaking, out-of-plane 
spin orientations are less common and typically weak~\cite{Yuan}.  As we discuss, 1T'-WTe$_2$ bucks this expectation: even though $E_\perp$ is out-of-plane, the non-aligned outer Te atoms (Fig.~\ref{fig1}c) enables an in-plane electric dipole to develop and a strong
out-of-plane 
spin orientation to be induced. 

Crucially, applied $E_\perp$ induces Berry curvature as well as magnetic moment. 
Berry curvature value is determined by an interplay between strong atomic (spin-selective) inter-orbital mixing of the 1T'-WTe$_2$ and $E_\perp$ induced terms, and exhibits 
a characteristic anisotropic distribution; magnetic moment mirrors this behavior (see Fig.~\ref{fig-s2}). While an electrically tunable Berry curvature can be readily realized in bilayer systems~\cite{Xiao07} owing to a electric control over layer degree of freedom, electrical tunability in {\it monolayer} systems is considerably more difficult. Berry curvature
is realizable in 1T'-WTe$_2$ as a direct result of the asymmetric non-aligned outer Te atoms.

Further, due to the low symmetry of 1T'-WTe$_2$ distorted crystal structure, induced Berry curvature and magnetic moment 
also possess an asymmetry characterized by a dipolar distribution in reciprocal space. As a result, shifts in the distribution function (e.g., induced when a dissipative charge current is flowing, $\vec j$) enable 
a net Berry flux, and a net out-of-plane magnetization, $M_z$, to develop (Fig.~\ref{fig4}). The latter corresponds to a direct (linear) magneto-electric effect 
$M_z = \sum_{i} \tilde{\alpha}_{z i} j_i$ ($i = x,y$),
where $\tilde{\alpha}_{z i}$ characterizes the strength of the magneto-electric effect; the former mediates a quantum nonlinear anomalous Hall effect~\cite{Sodemann}. 

Both these 
are intimately tied to the low-symmetry of gated 1T'-WTe$_2$; they do not appear in rotationally symmetric systems.
They constitute striking experimental signatures of the tunable quantum geometry (induced Berry curvature and magnetic moment) of 1T'-WTe$_2$, as well as the direct impact that its distorted structure has on its material response. 
Using available parameters for 1T'-WTe$_2$, we anticipate a sizeable $M_z$ that can be readily probed for e.g., using Kerr effect microscopy~\cite{lee2017}. 
1T'-WTe$_2$ provides a compelling venue to manipulate spins and magnetic moments in a tunable two-dimensional material. Out-of-plane spin orientations
are particularly useful since they may enable to couple to out-of-plane spins necessary for high-density magnetic applications~\cite{MacNeil,Kurebayashi}.

\vspace{2mm}

{\bf Symmetry analysis and $\vec k \cdot \vec p$ model ---} 
We begin by analyzing the band structure of monolayer 1T'-WTe$_2$ in the presence of an applied out-of-plane electric field $E_\perp$ [Fig.~\ref{fig1}(a)]. In doing so, we will employ a $\vec k \cdot \vec p$ method based on the underlying symmetries of the material: for e.g., mirror symmetry about the $xz$ mirror plane [dashed line in Fig.~\ref{fig1}(b)], time-reversal symmetry (TRS), and (broken) inversion symmetry (IS). For completeness, our analysis takes into account the three relevant atomic orbitals ($\psi_{1,2,3}$) and two spin states ($\uparrow,\downarrow$) that contribute to the states near the $\Gamma$ point and the gap opening [Fig.~\ref{fig1}(d)], as revealed by ARPES measurements~\cite{Tang} as well as first principles calculations~\cite{Qian, Choe, Lin,Xu}. This produces a six-band $\vec k \cdot \vec p$ description (SBD), see Appendix~\cite{Supp}, for a detailed account of the symmetry analysis of these orbitals and spin operators and the symmetry allowed terms in the SBD description. 

Importantly, while $E_\perp =0$ produces a spin-degenerate bandstructure~\cite{Tang, Qian, Choe, Lin,Xu}, 
when $E_{\perp}\neq0$ (e.g., induced by proximal gate) we find the bands become {\it spin-split} (Fig.~\ref{fig1}d). As shown in Fig.~\ref{fig1}d, this is particularly relevant away from the $\Gamma$ point, where the splitting becomes pronounced close to the band gap (gray shaded region, Fig.~\ref{fig1}d,e). These are characterized by states $\Psi_{\tau \xi}$ with higher ($\xi = +1$) or lower ($\xi = -1$) energies as shown in Fig.~\ref{fig1}e, and $\tau = \pm 1$ correspond to the conduction and valence bands.
As we will see, the splitting induced by $E_\perp$ drives a range of novel spin behavior. 

At low carrier densities typical for 1T'-WTe$_2$ devices~\cite{Wu}, the electronic and spin behavior is dominated by low-energy excitations around the bandgap in the four bands 
(Fig.~\ref{fig1}e). In order to compactly illustrate the physics, we develop a simple effective four-band model using the basis $\{ \psi_{c \uparrow}$, $\psi_{v \uparrow}$, $\psi_{c \downarrow}$, $\psi_{v \downarrow} \}$ in the regime around the gap opening (see gray region). This is obtained by performing a L\"owdin partitioning (see \cite{Supp}) of the bands in Fig.~\ref{fig1}d and can be expressed as 
$h^{\rm eff} (\vec k)= h_0(\vec k) + h_1 (\vec k)$. Here $h_0 (\vec k)$ describes the electronic behavior in pristine 1T'-WTe$_2$ ($E_\perp = 0$): 
\begin{align}
h_0(\vec k) = \bar{\epsilon}_{\vec k} + 
\lp
\begin{array}{cccc}
 m_{\vec k} & v^{+}_{\vec k} & 0 & 0 \\
 -v^-_{\vec k} & - m_{\vec k}  & 0 & 0 \\
 0 & 0 & m_{\vec k}  & v^{-}_{\vec k} \\
 0 & 0 & -v^{+}_{\vec k} & - m_\vec{k} 
\end{array}
\rp  ,
\label{eq:4band}
\end{align}
where $\bar{\epsilon}_{\vec k} = (\epsilon_{c \vec k} + \epsilon_{v \vec k})/ 2$, $m_{\vec k}  = (\epsilon_{c \vec k} - \epsilon_{v \vec k} ) / 2$, $v_{\vec k}^\pm  = \pm v_{x} k_x + i v_{y} k_y  $ represents the strong spin selective atomic orbital coupling (sharing the same spin), while $\epsilon_{c \vec k} $ and $\epsilon_{v \vec k} $ are diagonal parts for the conduction and valence bands, capturing their energy offsets and effective masses~\cite{Supp}. We note that $h_0(\vec k)$ is simply a {\it tilted} Bernevig-Hughes-Zhang (BHZ) hamiltonian~\cite{Bernevig} that describes the spin-degenerate bands in pristine 1T'-WTe$_2$; here the tilt arises from large effective mass differences between the conduction and valence bands (see Table.~\ref{tab:parameters}~\cite{Supp}).

On the other hand, $h_1(\vec k) = h_Z (\vec k) + h_R (\vec k) $ captures the electric field-induced spin-orbit coupling that are allowed by symmetry
\begin{align}
h_Z (\vec k) & = 
\begin{pmatrix}
\lambda k_y & i \delta_z & 0 & 0  \\
-i \delta_z & \lambda k_y & 0 & 0 \\
0 & 0 & -\lambda k_y & -i \delta_z  \\
0 & 0 & i \delta_z & -\lambda k_y 
\end{pmatrix} ,
\nn
h_R (\vec k) & = 
\begin{pmatrix}
0 & 0 & \alpha_{\vec k}^- & i \delta_x  \\
0 & 0 & i \delta_x & \alpha_{\vec k}^- \\
\alpha_{\vec k}^+ & -i \delta_x & 0 & 0  \\
-i \delta_x & \alpha_{\vec k}^+ & 0 & 0 
\end{pmatrix} ,
\label{eq:4bandSOC}
\end{align}
where we have grouped the electric field-induced spin-orbit coupling terms into $h_Z (\vec k)$ and $h_R(\vec k)$ in order to highlight the out-of-plane and in-plane spin orientations they induce respectively (see Fig.~\ref{fig2}). Here $\alpha_{\vec k}^\pm = \pm i \alpha_{x} k_x + \alpha_{y} k_y$, 
$\delta_{x,z}$ are $k$-independent coupling terms, and $\lambda k_y$ is a $k$-dependent that can induce out-of-plane spin orientation. We note, parenthetically, that the spin-orbit coupling terms in $h_Z$ have sometimes been referred to as ``Zeeman-like'' see e.g., Ref.~\cite{Yuan, Xu} so as highlight the out-of-plane spin orientation it induces; in the following, we will not use this terminology, but instead focus on their physical manifestation: its spin orientation. We emphasize that both $h_Z (\vec k)$ and $h_R(\vec k)$ in this work physically originate from IS breaking induced by the application of the electric field, see next section for detailed discussion. 

In writing Eq.~(\ref{eq:4bandSOC}) we have kept all symmetry allowed terms
up to the linear order in $k$ as allowed by symmetry. We remark that the magnitudes of each of the symmetry allowed terms can be determined from experimental or first principle calculation results (see Appendix~\cite{Supp} for a discussion). However, before we move to specific values, we first discuss the physical origin of the coupling terms, and some possible interplays brought by these couplings.

\vspace{2mm}

{\bf Physical origin of out-of-plane and in-plane spin orientations --- }
For physical clarity, we will denote $h_R (\vec k)$ and $h_Z (\vec k)$ as 
spin-orbit coupling induced by out-of-plane and in-plane IS breaking, respectively. In a general sense, $h_R (\vec k)$ [or $h_Z (\vec k)$] always couple states with opposite spins [the same spin]. As a result, these terms split the spin degeneracy and re-orient the spins
of the eigenstate $\Psi_{\tau \xi}$: $h_R$ terms create in-plane spin orientations (Fig.~\ref{fig2}a) whereas $h_Z$ terms align spins out-of-plane (Fig.~\ref{fig2}b). Here spin
orientations are plotted only for the right valley ($k_y>0$). For $k_y<0$, spin textures are flipped.

Their physical origins are also distinct. 
When a charge neutral monolayer 1T'-WTe$_2$ is placed under $E_\perp$, the top layer and the bottom layer Te atoms experience a charge redistribution becoming oppositely charged. This charge redistribution counteracts $E_\perp$ forming an out-of-plane dipole moment, $d_z$ (see Fig.~\ref{fig1}c). Crucially, because the Te atoms are not perfectly aligned, this charge redistribution also creates an {\it in-plane} electric dipole moment along the $x$-direction, $d_x$, yielding a net induced electric dipole, $\vec d_{\rm net}$, that is canted (see Fig.~\ref{fig1}c).

As a result, $E_\perp$ induces IS breaking in {\it both} $z$- and $x$-directions developing a nonzero $\p_z \phi (\vec r)$ and $\p_x \phi (\vec r)$; here $\phi(\vec r)$ is the local electro-static potential induced by $E_{\perp}$. Since spin-orbit coupling arises as matrix elements of the microscopic spin-orbit interaction: ${\hat H}_{\rm so} (\vec k) \sim ( \vec k + {\hat {\vec p}}/m_0 )\cdot \vec s \times \nabla \phi (\vec r)$~\footnote{see e.g., Eq.~(16.1) of Ref.~\cite{Bir}, or Eq.~(2.4) of Ref.~\cite{WinklerBook}}, we find that 
spin-orbit coupling terms $\delta_x$ and $\alpha_{\vec k}^\pm$ come from the total out-of-plane electric field $-\p_z \phi (\vec r)$; this constrains the terms $\vec k+{\hat {\vec p}}/m_0$ and $\vec s$ in ${\hat H}_{\rm so} (\vec k)$ to be in-plane only.
However, the charge re-distribution also enables an in-plane $\p_x \phi (\vec r)$ to develop. This in-plane electric field picks out $\vec s$ in ${\hat H}_{\rm so} (\vec k)$ as the out-of-plane component $s_z$, and $\vec k+{\hat {\vec p}}/m_0$ as the $y$-component. As a result of the in-plane $\p_x \phi (\vec r)$, $h_Z$ terms $\lambda k_y$ and $\delta_z$ in Eq.~\ref{eq:4bandSOC} manifest. 
The {\it distorted} structure of 1T'-WTe$_2$ enables $\vec d_{\rm net}$ that is generically canted with finite
$h_Z$ and $h_R$
terms that co-exist.
In contrast, since $\lambda k_y$ and $\delta_z$ result from the in-plane electric field $\p_x \phi (\vec r)$, the non-distorted  transition metal dichalcogenide monolayers whose atoms at top and bottom layers are aligned (e.g., MoS$_2$) do not possess an external $E_\perp$-induced out-of-plane spin orientations
near the $\Gamma$ point (up to linear in $k$). 

As a further illustration of the role that low symmetry plays in 1T'-WTe$_2$, we can also compare the spin-orbit coupling terms in $h_R(\vec k)$ allowed in 1T'-WTe$_2$ [induced by out-of-plane IS breaking], with that of HgTe quantum wells recently discussed in the literature~\cite{Tarasenko, Rothe}. 
For 1T'-WTe$_2$, all terms in $h_R (\vec k) $ are allowable when an out-of-plane electric field is applied because of its very low symmetry, and that fact that angular momentum in the $z$-direction is not a good quantum number.
In contrast, HgTe quantum wells possess a $D_{2d}$ symmetry, $[ h_R (\vec k) ]_{24,(42)}$ is missing because its 2nd and 4th basis functions are heavy hole bands with $j_z = \pm 3/2$ and the coupling between them is at least of $k^3$ order. Only $i \delta_x$ and $[ h_R (\vec k) ]_{13, (31)}$ can appear, corresponding to the existence of bulk inversion asymmetry \cite{Tarasenko} and structural inversion asymmetry \cite{Rothe}, respectively.

\vspace{2mm}

{\bf  Interplay between Berry curvature and the two types of spin-orbit couplings --- }
Pristine 1T'-WTe$_2$ possesses both IS and TRS ensuring that Berry curvature (and orbital magnetic moment) vanish exactly. As we now discuss, in 1T'-WTe$_2$ with $E_\perp \neq 0$, 
$h_Z$ [Eq.~(\ref{eq:4bandSOC})] presents an opportunity to break in-plane IS turning on a finite Berry curvature distribution. 
For clarity, we will first focus on the case when $\lambda$ and $\alpha_{x,y}$ are non-zero, while setting the $k$-independent terms $\delta_{x,z} = 0$, and then 
analyse the case when $\delta_{x,z} \neq 0$ later in the text.

To proceed, we first note that for pristine 1T'-WTe$_2$ (when $E_\perp = 0$), the hamiltonian $h^{\rm eff} = h_0 (\vec k)$ in Eq.~(\ref{eq:4band}) possesses spin degenerate states $\Psi_{\tau\xi}$, with $\xi = \pm 1$ corresponding to $\uparrow, \downarrow$ states, and 
spin-degenerate energy $\varepsilon_{\tau \xi} (\vec k) =  \bar{\epsilon}_{\vec k}  + \tau \Delta_{\vec k}$ where $\Delta_{\vec k}^2  = (v_x k_x )^2 + (v_y k_y )^2 + m_{\vec k}^2 $ is the energy difference between the conduction and valence bands. In the absence of $v_{x,y}$, conduction ($\psi_{\tau =+1,\xi}$) and valence ($\psi_{\tau =-1,\xi}$) bands touch, and exhibit a gapless spectrum along $\Gamma$-$Y$~\cite{Muechler}. However, large spin-selective atomic orbital coupling $v_{x,y}$ in 1T'-WTe$_2$ creates strong inter-orbital mixing (between $\psi_{c,v}$) giving a large QSH gap $\sim 0.055\, {\rm eV}$~\cite{Tang}.

Even though the external $E_{\perp}$ induced spin orbit coupling [Eq.~(\ref{eq:4bandSOC})] is small as compared with the intrinsic spin-selective atomic orbital coupling, $\alpha_{\pm}, \lambda \ll v_{\pm}$, nevertheless, when an external electric field $E_{\perp} \neq 0$ is applied, IS is immediately broken. Specifically, we emphasize that it is {\it in-plane} IS breaking that enables a finite Berry curvature, $\Omega_{\tau \xi} (\vec k)$, distribution to develop. As a result, we find that $\lambda$ encoding in-plane IS breaking (arising from $d_x$, Fig.~\ref{fig1}c) turns on $\Omega_{\tau \xi} (\vec k)$. In contrast, while $\alpha_{x,y}$ is also induced by $E_{\perp} \neq 0$ (and can also spin-split $\psi_{c,v}$ bands) it corresponds to an out-of-plane IS breaking, and does not lead to $\Omega_{\tau \xi} (\vec k)$.

\begin{figure}
\includegraphics[scale=0.21]{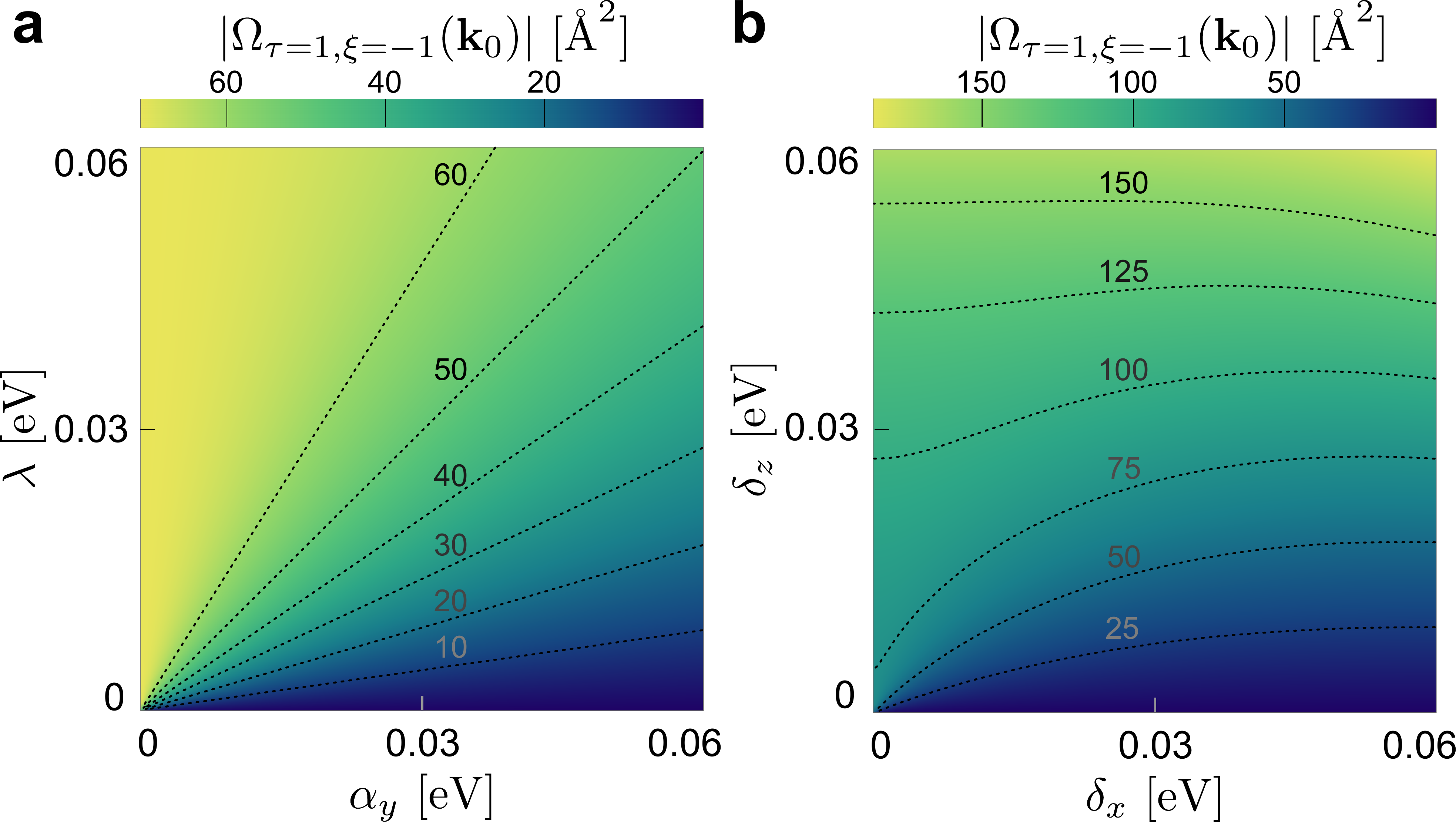}
\caption{
Peak Berry curvature value $\Omega_{\tau=1,\xi=-1} (\vec k_0)$ as a function of ({\bf a}) $k$-dependent spin-orbit coupling $(\alpha, \lambda)$ where we have set $\delta_{x,z}=0$, and ({\bf b}) $k$-independent spin-orbit coupling $(\delta_x, \delta_z)$ where we have set $\alpha,\lambda=0$. Peak values are taken at the band edge $\vec k_0= (0, k_0)$ with $k_0 = 0.385~{\rm \AA^{-1}}$.
Dashed lines denote equi-Berrry-curvature contours.
Both ({\bf a}) and ({\bf b}) show that Berry curvature is pronounced only when  
($\lambda$, or $\delta_z$) is 
large; this corresponds to strong in-plane inversion symmetry breaking.  
Parameters for the pristine part are listed in Table~\ref{tab:parameters}~\cite{Supp}.
}
\label{fig3}
\end{figure}

To see this explicitly, we first consider the case $\alpha \ll \lambda $ where
out-of-plane IS breaking is much weaker than in-plane IS breaking.
In this case,
$\lambda$ dominates $h_1(\vec k)$ and
we can take $\alpha \to 0$. Therefore, $h^{\rm eff}(\vec k) = h_0 (\vec k) + h_1(\vec k)$ produces a $\Psi_{\tau \xi}$ bandstructure with a lifted spin-degeneracy (Fig.~\ref{fig1}e) and 
energies $\varepsilon_{\tau \xi} (\vec k) = \bar{\epsilon}_{\vec k} + \tau \Delta_{\vec k} + \xi | \lambda k_y |$. We note that since both
$\lambda$ as well as spin-selective atomic orbital coupling $v_{x,y}$ do not mix spins, $\Psi_{\tau \xi}$ possess spins that purely point out of plane (Fig.~\ref{fig2}b). Using this, we find a Berry curvature distribution $\Omega_{\tau \xi} (\vec k) = \boldsymbol{\nabla}_{\vec k} \times \la  \Psi_{\tau \xi} (\vec k) |  i \boldsymbol{\nabla}_{\vec k} | \Psi_{\tau \xi} (\vec k) \ra$ as
\begin{align}
\Omega_{\tau \xi}^{(0)} (\vec k) = {\rm sgn}(k_y) \frac{\tau \xi}{2}  
\frac{v_{x} v_{y}}{\Delta_{\vec k}^3  }  (1-\vec k \cdot \nabla_{\vec k} ) m_{\vec k} .
\label{eq:omega0}
\end{align}
Strikingly, $\Omega_{\tau \xi}^{(0)} (\vec k)$ in Eq.~(\ref{eq:omega0}) does not depend on $\lambda$ even though finite
$\lambda$ was required to break in-plane IS. Instead, $\Omega_{\tau \xi}^{(0)} (\vec k)$ is solely determined by the spin-selective atomic orbital coupling $v_{x,y}$, and the band parameters in pristine 1T'-WTe$_2$. 

This decoupling behavior between IS breaking strength and the value of $\Omega_{\tau \xi}$ persists even in the presence of finite
out-of-plane IS breaking characterized by the ratios $\lambda/\alpha$.
To see this, we note that when $\alpha$ is finite, $h_1(\vec k)$ starts to hybridize $\Psi_{\tau \xi}$ with different spins (in the same $\tau$ band).
Since the intrinsic Berry curvature $\Omega_{\tau \xi}^{(0)} (\vec k)$ for spin up and spin down states are opposite in sign, when the $\xi$ states couple (via $\alpha$) the Berry curvature drops. 
Along the high symmetry line $k_x = 0$ about which the Berry curvature is even due to the TRS and mirror symmetry in the $y$-direction, the Berry curvature for the spin split bands near band edge can be expressed as \cite{Supp}
\begin{align}
\Omega_{\tau \xi} (k_x =0 ,k_y) = \frac{ \lambda
}{ (\lambda^2 + \alpha_y^2)^{1/2} } \Omega_{\tau \xi}^{(0)} (k_x =0 ,k_y),  
\label{eq:omega}
\end{align}
clearly displaying how $\Omega_{\tau \xi} (k_x =0 ,k_y)$ tends to the value expected in $\Omega_{\tau \xi}^{(0)}$ for $\lambda \gg \alpha$. In Fig.~\ref{fig3}a, we plot the peak value of Berry curvature $\Omega_{\tau=1,\xi=-1} $ reproduced in a numerical evaluation of the SBD description. This verifies our above analysis that the value of $\Omega_{\tau \xi}$ is bounded by the intrinsic (depends only on $v_{x,y}$) $\Omega_{\tau \xi}^{(0)} $,
and is tuned only by the ratios $\lambda/\alpha$, which makes equi-Berry-curvature contours to be straight lines (see Fig.~\ref{fig3}a).

We now consider the case of $\delta_{x,z} \neq 0$, while setting $k$-dependent terms $\alpha, \lambda = 0$. By numerically evaluating peak Berry curvature $\Omega_{\tau=1,\xi=-1} (\vec k_0)$
(see Fig.~\ref{fig3}b), we find that the peak Berry curvature $\Omega_{\tau=1,\xi=-1} (\vec k_0)$ develops a more complicated behavior. In particular, $\Omega_{\tau=1,\xi=-1} (\vec k_0)$  is no longer bounded by the intrinsic value $\Omega_{\tau \xi}^{(0)}$, and increases with $\delta_z$ without saturation. However, similar to the previous case, 
out-of-plane IS breaking $\delta_x$ alone is not able to induce a nonzero Berry curvature since it corresponds to an out-of-plane IS breaking. Large Berry curvature only appears when  
$\delta_z$ is significant. 

In the above, we concentrated on unveiling the (Berry curvature) features that the various symmetry allowed spin-orbit coupling terms possess. These features can in turn help to diagnose which of the ({\it a priori} symmetry-allowed) spin-orbit coupling terms dominate. We illustrate this by comparing with recent first principles calculations as well as a recent experiment \cite{Xu}.
In Ref.~\cite{Xu}, the Berry curvature of monolayer 1T'-WTe$_2$ was investigated at different perpendicular electric fields, from 0 to around 1~V nm$^{-1}$ using both first principles and a photocurrent measurement. In particular, their first principles results revealed the $E_\perp$ induced spin-splitting in the bandstructure that vanished at larger $k_y$ away from the band edge, and a peak Berry curvature that increased with $E_\perp$. This observation means that $k$-dependent spin-orbit couplings may play only a minimal role. Further both the experiment and the first principles calculations found large Berry curvature at band edge even at small electric field shows that $\delta_z > \delta_x$ (see Fig.~\ref{fig3}). Strikingly, these values of Berry curvature are close to the large intrinsic values expected from $v_{x,y}$ and $m_{\vec k}$. Together with Fig.~\ref{fig3}b 
this indicates that the 
in-plane IS breaking and $h_Z$ terms 
dominates, overwhelming the $h_R$ terms. As a result, in what follows we will use the k-independent $\delta_z$ spin-orbit coupling term to describe the $E_\perp$ induced spin texture.

\vspace{2mm}

\begin{figure*}[t!]
\includegraphics[scale=0.22]{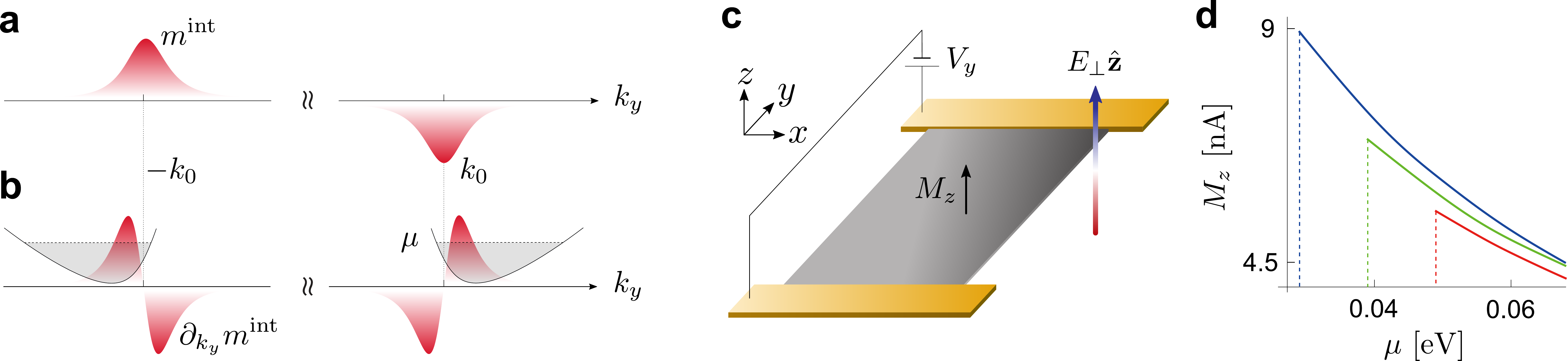}
\caption{
({\bf a},{\bf b}) Asymmetry in 1T'-WTe$_2$ enable dipolar distributions of 
intrinsic magnetic moment ({\bf a}) $m^{\rm int}$
characterized by a 
({\bf b}) $\p_{k_y} m^{\rm int}$ 
near the gap opening $\pm \vec k_0 = (0, \pm k_0)$. Black curves represent the conduction band ($\tau = 1, \xi=-1$), dashed lines denote the chemical potential $\mu$ with gray shaded areas are the occupied states.
Due to the tilt of the band dispersion, 
an intrinsic magnetic moment dipole arises near the gap opening, and gives rise to 
a non-zero current-induced magnetization $M_z$ [see Eq.~(\ref{eq:MzJ})]. 
Parameters used for the pristine part are the same as those in Fig.~\ref{fig1}
for an illustration.
({\bf c}) Schematic of a 1T'-WTe$_2$ monolayer under a perpendicular electric field $E_\perp \hat{\vec z}$ and an in-plane electric field $E_y$, which can give rise to intrinsic magnetization $M_z$.
({\bf d}) Calculated intrinsic magnetization $M_z$ 
induced by a current $j_y = 10~{\rm A\, m^{-1}}$ in the $y$-direction from Eq.~(\ref{eq:MzJ}), 
with $\lambda,\alpha,\delta_x =0$, 
but varying $\delta_z = 0.075~{\rm eV}$ (blue), $0.05~{\rm eV}$ (green), and $0.025~{\rm eV}$ (red).
These magnitudes for $\delta_z$ are within reach by applying an $E_\perp < 1~\rm{V \, nm}^{-1}$~\cite{Xu}.
The $M_z$ plotted here becomes non-zero only when the chemical potential is above the conduction band bottom (denoted by dashed lines), 
which is different for each of the lines.
The trend from the red line to the blue line shows that a larger 
$\delta_z$ (stronger in-plane IS breaking)
leads to an increased intrinsic magnetic moment as well as more pronounced $M_z$, as illustrated in Fig.~\ref{fig2}.}
\label{fig4}
\end{figure*}

{\bf Current induced magnetization ---}
Another closely related quantity, the (intrinsic) orbital magnetic moment $m_n^{\rm int} (\vec k)$, also appears when in-plane IS is broken by $E_\perp \hat{\vec{z}}$:
\begin{align}
m_{n}^{\rm int} (\vec k) = 
\frac{e}{\hbar} \Re \sum_{n' \neq n}
\frac{i \la n | \p H / \p k_x  | n' \ra \la n' | \p H / \p k_y  | n \ra }
{ \varepsilon_n - \varepsilon_{n'} },
\label{eq:omm}
\end{align}
where we have written $n = \{\tau \xi\}$ as a short-hand, and $H$ is the hamiltonian. The orbital magnetic moment comes from the self-rotation of a Bloch electron wave packet around its center of mass, and is an intrinsic property of the Bloch band~\cite{Xiao}, and its distribution in momentum space mimics that of the Berry curvature distribution (see Appendix~\cite{Supp}). 

The low-symmetry of 1T'-WTe$_2$ enables an asymmetric distribution of of $\Omega_n (\vec k)$ and $m_n (\vec k)$ (see Fig.~\ref{fig4} a,b).
This affords the opportunity to realize Berry phase effects not normally achievable in their rotational symmetric cousins.
A striking example is the (linear) magneto-electric effect (ME) $M_{z} = \sum_{i} \alpha_{z i} E_i$ ($i = x,y$), where the flow of an in-plane current induces an out-of-plane magnetization. While typically found in multi-ferroic materials~\cite{Eerenstein} where TRS and IS are explicitly broken, ME effects can arise in metals with sufficiently low symmetry (broken IS as well as broken rotational symmetry) and where the dissipation of a charge current breaks TRS~\cite{Levitov}.  
\footnote{
We can see these from the symmetry analysis as follows: (a) in a time-reversal symmetric system 
$M_z = \sum_{i} \alpha_{z i} E_i$, then under the symmetry operation $t \to -t$ we have $M_z \to - M_z$, $\vec E \to \vec E$, and $M_z = \sum_{i} \alpha_{z i} E_i \to - M_z = \sum_{i} \alpha_{z i} E_i$ which leads to $\alpha_{z x} = \alpha_{z y} = 0$;
(b) if a system has a rotational symmetry,  
then under the rotation by angle $\theta$ 
we have $M_z \to M_z$, $E_x \to E_x \cos \theta - E_y \sin \theta$, $E_y \to E \sin \theta + E_y \cos \theta$, and 
$M_z = \alpha_{zx} E_x + \alpha_{zy} E_y \to M_z = (\alpha_{zx} E_x + \alpha_{zy} E_y) \cos \theta + (\alpha_{zy} E_x - \alpha_{zx} E_y) \sin \theta$ which leads to $\theta = 2 n \pi$ ($n = 0,1,2, \dots$), i.~e., rotational symmetry is not allowed for non-zero $\alpha_{zx}$ and $\alpha_{zy}$;
(c) if a system is centrosymmetric, then under the
operation $(x,y) \to (-x, -y)$ we have $M_z \to M_z$, $\vec E \to -\vec E$, and $M_z = \sum_{i} \alpha_{z i} E_i \to M_z = - \sum_{i} \alpha_{z i} E_i$ which also requires $\alpha_{z x} = \alpha_{z y} = 0$.
}. This is termed the kinetic ME effect~\cite{Levitov,Sahin}.

Indeed, the low-symmetry of 1T'-WTe$_2$ (with $E_\perp \neq 0$) where only a mirror symmetry in the $y$-direction remains is the largest symmetry group that hosts the kinetic ME~\cite{Levitov, Sodemann}. This makes 1T'-WTe$_2$ a natural venue to control ME.

To illustrate the kinetic ME effect in 1T'-WTe$_2$, we first note that the magnetic moment is asymmetric, displaying a dipolar distribution (see Fig.~\ref{fig4}a,b). This can be seen explicitly by considering $\partial_{k_y} m$ and noting that it is displaced in relation to the bottom of the band, Fig.~\ref{fig4}b. As a result, when an in-plane electric field shifts the distribution function, a uniform out-of-plane magnetization $M_z$ develops:
\begin{align}
M_z
 = \sum_{i=x,y} \tilde{\alpha}_{zi} j_i , \quad 
\tilde{\alpha}_{zi} = \Big[\frac{e}{\hbar}  \sum_{n,\vec k} 
f_{n \vec k}^{(0)}
\frac{\p m_n^{\rm tot} (\vec k) }{\p k_i } \Big] (D_{ii})^{-1}   
\label{eq:MzJ}
\end{align}
where $D_{ii}$ is the Drude weight along $i$ direction, $f_{n \vec k}^{(0)}$ is the equilibrium distribution function, 
$m_n^{\rm tot} (\vec k) = m_n^{\rm int} (\vec k) + (g e/2 m_0) \la u_{\vec k} | s_z | u_{\vec k} \ra$ is the intrinsic contribution to the magnetic moment in a particular band, containing both orbital and spin contributions with $s_z = \hbar \sigma_z / 2$. For 1T'-WTe$_2$ monolayer, we estimate $g \sim 5$~\cite{Wu,Fei}.

Importantly, Eq.~(\ref{eq:MzJ}) reflects the symmetry of the crystal. For example, magnetic moment distribution has equal magnitudes but opposite signs in the two electron pockets in the conduction band. As a result, $\tilde{\alpha}_{zx} =0$ vanishes as expected from symmetry, see above. In contrast, when in-plane electric field is applied along $y$, a non-zero $M_z$ is generated (i.e. $\tilde{\alpha}_{zy} \neq 0$).

Using Eq.~(\ref{eq:MzJ}),
we obtain a finite out-of-plane magnetization $M_z$ in Fig.~\ref{fig4}d 
when current is driven along the $y$ direction. In doing so, we used $f_{n \vec k}^{(0)} = \Theta(\varepsilon_{n \vec k} - \mu)$ with $\mu$ the chemical potential, and computed the Drude weight in the usual fashion. Further, to capture the full reciprocal space distribution of the magnetic moment (including regions away from the gap opening), we used the SBD description to compute the magnetic moment distribution. 
Here we have concentrated on small chemical potentials so that only moments in the lowest conduction band (blue band in Fig.~\ref{fig1}e) contribute. Since $m_n^{\rm int} (\vec k)$ is an odd function of $k_y$ when TRS is present, the filled bands do not contribute to ME. This reflects the fact that kinetic ME arises from a dissipative process. As a result, when the chemical potential is in the gap, $\tilde{\alpha}_{zy} =0$. However, once the system is doped into the conduction band, a non-zero ME develops, see Fig.~\ref{fig4}d. 
Similar analysis also applies to the Berry curvature $\Omega_{\tau \xi}$ (which exhibits a dipolar distribution), and leads to a non-linear Hall effect without applied magnetic field (see Ref.~\cite{Sodemann} as well as the Appendix~\cite{Supp} 
for an explicit discussion for this system).

\vspace{2mm}
{\it Summary --} 1T'-WTe$_2$ with an applied out-of-plane electric field $E_\perp$, provides a new and compelling venue to control bulk band quantum geometry. In particular, its bands exhibit a tunable Berry curvature and magnetic moment with switch-like behavior. Crucially, the low symmetry of its crystal structure enable effects not normally found in its rotationally symmetric cousins. These include striking Berry phase effects such as a current induced magnetization (ME), and a quantum non-linear Hall effect.
These are particularly sensitive to orientation of in-plane electric field and the crystallographic directions. Indeed, $M_z$ is strongest when current runs along the $y$-direction; this sensitivity can be verified through measurements in a single 1T'-WTe$_2$ sample, for e.g., using a Corbino disc geometry.  Perhaps most exciting, however, is how IS broken 1T'-WTe$_2$ enables direct and electric-field tunable access to out-of-plane magnetic degrees of freedom. Given its two-dimensional nature, 1T'-WTe$_2$ can be stacked with other two-dimensional materials, providing a key magneto-electric component in creating magnetic van der Waals heterostructures.

\vspace{2mm}
{\bf Acknowledgements} - We gratefully acknowledge useful conversations with Valla Fatemi, Qiong Ma, Su-Yang Xu, and Dima Pesin. This work was supported by the Singapore National Research Foundation (NRF) under NRF fellowship award NRF-NRFF2016-05 and a Nanyang Technological University Start-up grant (NTU-SUG).


\clearpage

\newpage

\setcounter{equation}{0}
\renewcommand{\theequation}{A-\arabic{equation}}
\renewcommand{\thefigure}{A-\arabic{figure}}
\renewcommand{\thetable}{A-\Roman{table}}
\makeatletter
\renewcommand\@biblabel[1]{A#1.}
\setcounter{figure}{0}
\twocolumngrid


\section*{Appendix for ``Symmetry, spin-texture, and tunable quantum geometry in WTe$_2$ monolayer''}

\subsection{Six-band $\vec k \cdot \vec p$ model for monolayer 1T'-WTe$_2$}

Without external fields, 1T'-WTe$_2$ monolayers possess time-reversal (TR) symmetry and a point symmetry group $P2_{1}/m$ that contains four symmetry operations. If we set one of the inversion center as the origin ${\cal O}_0 = (0,0,0)$ in real space, the four symmetry operations are
$\{ (x,y,z), (-x,-y,-z), (x,1/2-y,z), (-x,1/2+y,-z) \}$ where $1/2$ denotes shifting by $1/2$ a unit cell in the $y$-direction. When we shift the origin ${\cal O}_0$
to ${\cal O}_1 = (0,1/2,0)$, the four symmetry operations become
$\{ (x,y,z), (x,-y, z), (-x,1/2,-z), (-x,-1/2,-z) \}$.

Various first-principle calculations as well as expermental measurements~\cite{Qian,Tang,Choe,Lin} showed that there are three relevant orbitals contributing to the states near the gap. Although the exact orbital compositions at the $\Gamma$ point is not clear, these orbitals at the $\Gamma$ point are consistently revealed~\cite{Tang,Choe,Lin} to be (even, odd, even) under the reflection operation in $y$-direction (the orbitals are ordered with decreasing energy). Here we note that Choe {\it et al.} used a different coordinate system with $x$ and $y$ directions exchanged.

A perpendicularly applied electric field breaks symmetry operations that flip in the $z$-direction and reduce the point group $P2_1/m$ to group $C_{1v}$ that has only two symmetry operations $\{\mathbb{I}, M_y \}$ (i.~e., identity and reflection about the $xz$ plane that cross a Te atom). It has two real 1D irreducible representations (see Table \ref{supptab:characterG}).

Aside from the spin degrees of freedom, each of the three orbitals at the $\Gamma$ point is non-degenerate and transforms according to one of the 1D irreducible representations of $C_{1v}$. Moreover, although inversion symmetry is broken, the three orbitals remain (even, odd, even) in the $y$-direction at the $\Gamma$ point. These two observations show that the three orbitals at the $\Gamma$ point transform as:
\begin{align}
\psi_1 \sim 1 ,
\quad
\psi_2 \sim y  ,
\quad
\psi_3 \sim 1 ,
\label{suppeq:basis}
\end{align}
where the symbol ``$\sim$'' denotes how these functions transform under operations in $C_{1v}$.
Using $\{ \psi_1$, $\psi_2$, $\psi_3\}$ as basis, the $\vec k \cdot \vec p$ Hamiltonian near the $\Gamma$ point assumes a $3 \times 3$ form:
\begin{align}
H (\boldsymbol{\cal K})  =
\lp
\begin{array}{ccc}
H^{11} (\boldsymbol{\cal K})& H^{12} (\boldsymbol{\cal K})& H^{13}(\boldsymbol{\cal K})\\
H^{21} (\boldsymbol{\cal K})& H^{22} (\boldsymbol{\cal K})& H^{23}(\boldsymbol{\cal K})\\
H^{31} (\boldsymbol{\cal K})& H^{32} (\boldsymbol{\cal K})& H^{33}(\boldsymbol{\cal K})\\ 
\end{array}
\rp  ,
\end{align}
where $H^{\alpha \beta} (\boldsymbol{\cal K}) $ is the $2 \times 2$ ($1 \times 1$) block matrix between $\psi_1$ and $\psi_2$ with (without) spin degree of freedom included.

In the following, we will obtain the general form of $H (\boldsymbol{\cal K})$ from symmetry analysis. The necessary information for $\psi_{1,2,3} $ is contained in its transformation property under group $C_{1v}$ [Eq.~(\ref{suppeq:basis})]. We note that detailed orbital compositions, e.~g., the weight of $p$- or $d$- orbital in $|\psi_{1,2,3} \ra $, does not affect the following analysis.

We will proceed by using the theory of invariants~\cite{Bir,Winkler,Zhou} which is based on the invariance of the Hamiltonian $\hat H$ under all operations of the corresponding crystal symmetry group. When the Hamiltonian $\hat H$ is projected to  the energy bands of interest $\hat H = \sum_{\alpha,\beta} | \psi_\alpha \ra  H^{\alpha \beta} (\boldsymbol{\cal K})  \la \psi_\beta  | $, where $\boldsymbol{\cal K}$ denotes a tensor operator formed by combinations of  wave vectors, the symmetry group constrains the $\vec k \cdot \vec p$ Hamiltonian $H^{\alpha \beta} (\boldsymbol{\cal K})$ as follows: under an arbitrary symmetry operation $g \in {C_{1v}}$, the basis $| \psi_\alpha \ra$ transforms according to the irreducible representation $\Gamma_\alpha$, so the invariance of the Hamiltonian under the symmetry operation $g$ dictates
$\hat{P}_g {\hat H} \hat{P}_g^{-1}  = {\hat H}$, where $\hat{P}_g$ denotes the operator for symmetry operation $g$. This leads to $\sum_{\alpha', \beta'} \la \psi_\alpha  | \hat{P}_g | \psi_{\alpha'} \ra H^{\alpha' \beta'} (\hat{P}_g  \boldsymbol{\cal K}  \hat{P}_g^{-1} ) \la \psi_{\beta'}  |  \hat{P}_g^{-1}  | \psi_\beta \ra   = H^{\alpha \beta} ( {\boldsymbol{\cal K}}  )$, or equivalently,
\begin{align}
\vec D^{\alpha} (g) H^{\alpha \beta} ( \hat{P}_g {\boldsymbol{\cal K}} \hat{P}_g^{-1} ) \vec D^{\beta} (g^{-1})\ = H^{\alpha \beta} ( {\boldsymbol{\cal K}}  ) ,
\label{suppeq:invarient}
\end{align}
where $\la \psi_\alpha  | \hat{P}_g | \psi_{\alpha'} \ra = \delta_{\alpha \alpha'} \vec D^{\alpha} (g)$ with $\vec D^{\alpha} (g)$ is the representation matrix of $g$ in $\Gamma_\alpha$ (in the 1D irreducible representation case here, $1$ or $-1$), and $\hat{P}_g {\boldsymbol{\cal K}} \hat{P}_g^{-1}$ denotes the transformation of ${\boldsymbol{\cal K}}$ under the symmtery operation $g$, e.~g., if $g = M_y$ and ${\boldsymbol{\cal K}} = k_y$, then $\hat{P}_{M_y} k_y \hat{P}_{M_y}^{-1} = -k_y$. 

In constructing the $\boldsymbol{\cal K}$ operators, one can also take into account the spin degree of freedom, by including the spin operator $\vec s = (s_x , s_y , s_z)$ in the $\vec k \cdot \vec p$ Hamiltonian. Note that $\vec s$ is a pseudovector, we have $s_x \to - s_x$, $s_z \to - s_z$, and $s_y \to s_y$ under the operation $M_y$~\cite{Zhou}, e.~g., if $g = M_y$ and ${\boldsymbol{\cal K}} = k_y s_x$, then $\hat{P}_{M_y} k_y s_x \hat{P}_{M_y}^{-1} = ( \hat{P}_{M_y} k_y \hat{P}_{M_y}^{-1} ) ( \hat{P}_{M_y} s_x \hat{P}_{M_y}^{-1} ) = (-k_y)(-s_x) = k_y s_x$.

\vspace{1mm}
For general cases in which crystals have high symmetry point groups, the expression of an arbitrary block $H^{\alpha \beta} ( {\boldsymbol{\cal K}} )$ of the $\vec k \cdot \vec p$ Hamiltonian can be constructed in several standard procedures with the corresponding full character table~\cite{Bir,Winkler,Zhou}. In our case, however, the group $C_{1v}$ is the simplest non-trivial group that has only two symmetry operations $\{\mathbb{I}, M_y \}$, and we can do the analysis just based on mirror symmetry operation $M_y$:

i) For blocks $H^{\alpha \alpha}(\boldsymbol{\cal K})$ ($\alpha = 1,2,3$) and $H^{1 3}(\boldsymbol{\cal K})$: since $| \psi_\alpha \ra \la \psi_\alpha  |$  ($\alpha = 1,2,3$) and $| \psi_1 \ra \la \psi_3  |$ are even under $M_y$ operation, to make sure $\hat H$ is invariant under $M_y$ operation, then $H^{\alpha \alpha} (\boldsymbol{\cal K})$ ($\alpha = 1,2,3$) and $H^{13} (\boldsymbol{\cal K})$ must be composed by $\boldsymbol{\cal K}$ operators that are also even under $M_y$ operation. Relevant operator combinations that are invariant under $M_y$ operation are listed in the first row of Table \ref{supptab:characterG}.
\begin{table}[h!]
\centering
\caption{The character table for the group $C_{1v}$, and corresponding operator combinations that are even (first row) and odd (second row) under $M_y$. $s_0$ is the $2 \times 2$ identity matrix. Note that the prefactor $i$ in some of the terms guarantees TR symmetry, since $i$, $\vec k$, and $\vec s$ are odd under the TR operation. Here we only keep terms up to $O(k^2)$ in the diagonal part and $O(k)$ in the off-diagonal part.}
\label{supptab:characterG}
\hspace{0cm}
\begin{tabular}{cccc}
\hline\hline
    $C_{1v}$  \qquad \qquad& $\mathbb{I}$ & $M_y$ & \qquad TR invariant operators \\
    \hline
    $\Gamma_1$ \qquad \qquad&  $ 1$ & $ 1$  & \qquad $s_0$, $i k_x$, $k_x^2$, $k_y^2$, $i s_y$, $k_x s_y$, $k_y s_x$, $k_y s_z$ \\
    $\Gamma_2$ \qquad \qquad&  $ 1$ & $-1$  & \qquad $i k_y$, $i s_x$, $i s_z$, $k_x s_x$, $k_x s_z$, $k_y s_y$  \\
    \hline\hline
\end{tabular}
\end{table}

ii) For blocks $H^{12}(\boldsymbol{\cal K})$ and $H^{23}(\boldsymbol{\cal K})$: since $| \psi_1 \ra \la \psi_2  |$ and $| \psi_2 \ra \la \psi_3  |$ are odd under $M_y$ operation, then $H^{12} (\boldsymbol{\cal K})$ and $H^{23} (\boldsymbol{\cal K})$ must be composed by $\boldsymbol{\cal K}$ operators that are also odd under $M_y$ operation to ensure that $\hat H$ is invariant under $M_y$ operation. Relevant operator combinations that are odd under $M_y$ are listed in the second row of Table \ref{supptab:characterG}.

After obtaining the terms that transform correctly for each of the blocks, we note further constraints that trim the $\vec k \cdot \vec p$ Hamiltonian:

(1) The $\vec k \cdot \vec p$ Hamiltonian must be Hermitian, which dictates that the invariants for a diagonal block must be Hermitian.

(2) Terms containing both $\vec k$ and $\vec s$ are matrix elements of the microscopic spin-orbit interaction ${\hat H}_{\rm so} (\vec k) \sim \vec k \cdot [\vec s \times \nabla V(\vec r)]$ in $\hat H$ thus terms $k_x s_x$ and $k_y s_y$ will not appear.

(3) For our purposes of estimating the Berry curvature and orbital magnetic moments in the main text, we can neglect the $H^{13}$ block. This is because $| \psi_1 \ra$ and $| \psi_3 \ra$ are energetically far away from each other ($\gtrsim 0.5$  eV) and their couplings only have small contributions to Berry curvature and orbital magnetic moment for the conduction bands. Although $H^{13}$ block contributes to optical transitions $\gtrsim 0.5$  eV, this is beyond our current scope.

The above three considerations trim/eliminate the terms $\{ i k_x, i s_y, k_x s_x, k_y s_y \}$. For remaining terms, we group them into terms induced by the applied perpendicular electric field, and those that are present in pristine 1T'-WTe$_2$. To do so we perform the following symmetry analysis: when the electric field is not present, inversion symmetry about the inversion center ${\cal O}_0$ is recovered, and the operation $(-x, 1/2 ,-z)$ about ${\cal O}_1$ becomes a symmetry operation again. Under the operation $(-x, 1/2 ,-z)$, there is no flip in the $y$-direction while $(k_x, k_y) \to (-k_x, k_y)$ and $(s_x, s_y, s_z) \to (- s_x, s_y, -s_z)$. We can see that terms $\{ i s_x, i s_z, k_x s_y, k_y s_x, k_y s_z \}$ change sign under this new operation, i.~e., they are not invariant under the original symmetry group $P2_{1}/m$ and can appear only when the perpendicular electric field is applied.
\begin{table}[b]
\centering
\caption{Terms that can appear for $\vec k \cdot \vec p$ Hamiltonian blocks. The prefactor $i$ in some of the terms guarantees TR symmetry, since $i$, $\vec k$, and $\vec s$ are odd under the TR operation. $s_0$ is the $2 \times 2$ identity matrix. Similar to above, we kept terms up to $O(k^2)$ in the diagonal part and $O(k)$ in the off-diagonal part.}
\label{supptab:invarients}
\hspace{0cm}
\begin{tabular}{cccc}
\hline\hline
   \qquad \qquad & $H^{\alpha \alpha} (\boldsymbol{\cal K})$   \qquad \qquad  & $H^{12} (\boldsymbol{\cal K})$, $H^{23} (\boldsymbol{\cal K})$ \\
\hline
with or w/o $E{\hat z}$ \qquad \qquad & $s_0$, $k_x^2 s_0$, $k_y^2 s_0$  \qquad \qquad &  $i k_y s_0$, $k_x s_z$ \\
\hline
with $E{\hat z}$  \qquad \qquad &   $k_y s_z$; $k_x s_y$, $k_y s_x$   \qquad \qquad &  $i s_z$; $i s_x$ \\
\hline\hline
\end{tabular}
\end{table}

After trimming, classification, and analyzing the physical origin of the terms induced by electric field, we now obtain the general form of the $\vec k \cdot \vec p$ Hamiltonian (see Table~\ref{supptab:invarients}).

\begin{figure}[t]
\includegraphics[width=0.68\columnwidth]{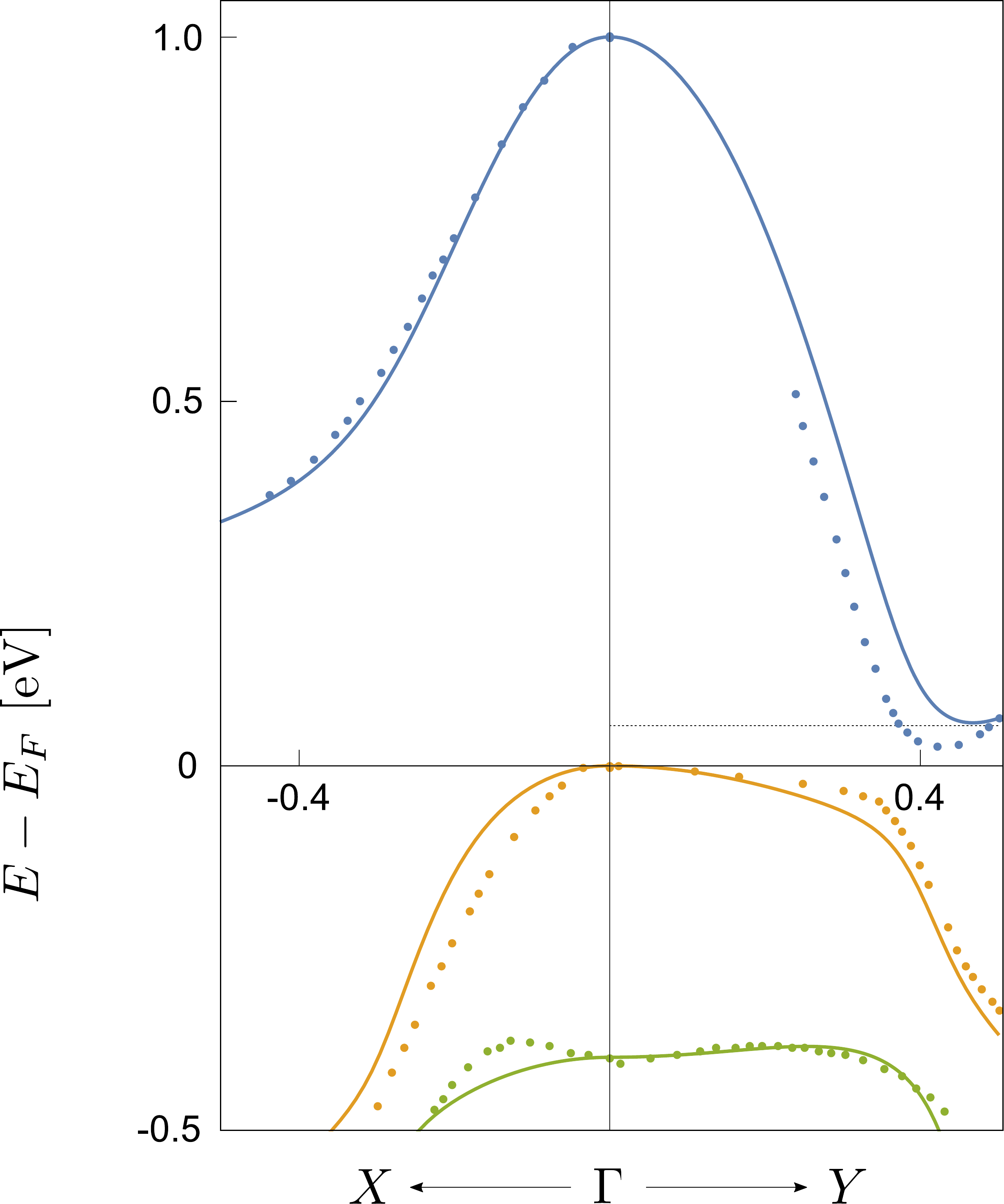}
\caption{Energy dispersion near the Fermi surface from the model Eq.~(\ref{suppeq:H0}), with an overall band gap 0.055 eV~\cite{Tang}, indicated by the dashed line)  (solid lines). Data (dots) extracted from Ref.~\cite{Qian,Xu}. The parameters for the model are listed in Table~\ref{tab:parameters}.}
\label{fig-s1}
\end{figure}

First we use a least squares fitting to extract coefficients of these invariant operators from known band structure, either from experimental measurements or numerical calculations. From the first principle calculation result~\cite{Qian, Xu}, we obtained the Hamiltonian $H_0 (\vec k)$ when there is no external fields. We find 
\begin{align}
H_0 (\vec k) =
\lp
\begin{array}{cccccc}
\epsilon_1 & 0 & v_1^+ & 0 & 0 & 0 \\
0 & \epsilon_1 & 0 & v_1^- & 0 & 0\\
-v_1^- & 0 & \epsilon_2 & 0 & v_3^+ & 0\\
0 & -v_1^+ & 0 & \epsilon_2 & 0 & v_3^-\\
0 & 0 & -v_3^- & 0 & \epsilon_3 & 0\\
0 & 0 & 0 & -v_3^+ & 0 & \epsilon_3
\end{array}
\rp  .
\label{suppeq:H0}
\end{align}
where $\epsilon_i = c_{i,0} + c_{i,x} k_x^2 + c_{i,y} k_y^2$, and $v_i^\pm = \pm v_{i, x} k_x + i v_{i, y} k_y  $. The dispersion is plotted in Fig.~\ref{fig-s1}, with parameters listed in Table~\ref{tab:parameters}.
\begin{table}[h!]
\centering
\caption{Least squares fitted $\vec k \cdot \vec p$ parameters for a 1T'-WTe$_2$ monolayer without external fields from Ref.~\cite{Qian,Xu}, with an overall band gap 0.055 eV.}
\label{tab:parameters}
\hspace{0cm}
\begin{tabular}{cccccc}
\hline\hline
Parameter  \qquad & Value  \qquad & Unit  \qquad \qquad & Parameter  \qquad & Value  \qquad & Unit  \\
\hline
$c_{1,0}$  \qquad & $1.0$  \qquad & eV  \qquad \qquad &   \qquad &   \qquad &   \\
$c_{2,0}$  \qquad &    $0$  \qquad & eV  \qquad \qquad &   \qquad &   \qquad &   \\
$c_{3,0}$  \qquad & $-0.4$  \qquad & eV  \qquad \qquad &   \qquad &   \qquad &  \\
\hline
$c_{1,x}$  \qquad & $-11.25$  \qquad & eV {\AA}$^2$  \qquad \qquad & $c_{1,y}$  \qquad & $-6.90$  \qquad & eV {\AA}$^2$  \\
$c_{2,x}$  \qquad & $-0.27$  \qquad & eV {\AA}$^2$  \qquad \qquad & $c_{2,y}$  \qquad & $-1.08$  \qquad & eV {\AA}$^2$  \\
$c_{3,x}$  \qquad & $-0.82$  \qquad & eV {\AA}$^2$  \qquad \qquad & $c_{3,y}$  \qquad & $0.99$  \qquad & eV {\AA}$^2$  \\
\hline
$v_{1, x}$  \qquad & $1.71$  \qquad & eV {\AA}  \qquad \qquad & $v_{1, y}$  \qquad & $0.48$  \qquad & eV {\AA}  \\
$v_{3, x}$  \qquad & $0.48$  \qquad & eV {\AA}  \qquad \qquad & $v_{3, y}$  \qquad & $-0.48$  \qquad & eV {\AA}  \\
\hline\hline
\end{tabular}
\end{table}

The additional terms that are induced by the applied perpendicular electric field $E_\perp$ makes
the full Hamiltonian $H (\vec k) = H_0 (\vec k) + H_1 (\vec k)$, with
\begin{align}
H_1 (\vec k)  =
\lp
\begin{array}{cccccc}
\lambda_1 k_y & \alpha_1^- & i \delta_{1, z} & i \delta_{1, x} & 0 & 0 \\
\alpha_1^+ & -\lambda_1 k_y & i \delta_{1, x} & -i \delta_{1, z} & 0 & 0\\
-i \delta_{1, z} & -i \delta_{1, x} & \lambda_2 k_y & \alpha_2^- & i \delta_{3, z} & i \delta_{3, x} \\
-i \delta_{1, x} & i \delta_{1, z} & \alpha_2^+ & -\lambda_2 k_y & i \delta_{3, x} & -i \delta_{3, z} \\
0 & 0 & -i \delta_{3, z} & -i \delta_{3, x} & \lambda_3 k_y & \alpha_3^-\\
0 & 0 & -i \delta_{3, x} & i \delta_{3, z} & \alpha_3^+ & -\lambda_3 k_y
\end{array}
\rp  ,
\label{suppeq:H1}
\end{align}
where
$\alpha_i^\pm = \pm i \alpha_{i, x} k_x + \alpha_{i, y} k_y$ is the 
commonly seen 
spin-orbit coupling for $i$-th orbital (this is sometimes referred to as ``Rashba'' spin texture), 
$\lambda_i k_y$ is a 
spin splitting from in-plane IS breaking,
and $\delta_{i, x}$ and $\delta_{i, z}$ are $k$-independent inter-band couplings, see main text for discussion of physical origin. 


\subsection{$4 \times 4$ model near the band gap}

To obtain the effective $4 \times 4$ Hamiltonian near the band gap, we consider the energy eigenvalue equation
\begin{align}
\lp
\begin{array}{cc}
h_q & u \\
u^\dagger & h_d	 
\end{array}
\rp
\lp
\begin{array}{c}
\psi_q \\
\psi_d 	 
\end{array}
\rp
=
\varepsilon
\lp
\begin{array}{c}
\psi_q \\
\psi_d 	 
\end{array}
\rp  ,
\label{suppeq:eigen}
\end{align}
where $h_q$ ($h_d$) is the $4 \times 4$ ($2 \times 2$) diagonal block from the original $6 \times 6$ Hamiltonian, and $\psi_q$ ($\psi_d$) is the corresponding four (two) component state vector, and $u$ is the $4 \times 2$ matrix couples $h_q$ and $h_d$.

The second row of Eq.~(\ref{suppeq:eigen}) allows $\psi_d$ to be written in terms of $\psi_q$:
\begin{align}
\psi_d = (\varepsilon - h_d)^{-1} u^\dagger \psi_q  .
\end{align}
Substituting this into the first row of Eq.~(\ref{suppeq:eigen}) gives an effective eigen equation solely for the $\psi_q$ components:
\begin{align}
\lb h_q + u (\varepsilon - h_d)^{-1} u^\dagger  \rb   \psi_q = \varepsilon \psi_q  .
\end{align}
Performing the standard expansion in small $\varepsilon$ as well as rotation procedure~\cite{Winkler}, we obtain the effective Hamiltonian near the band gap as
\begin{align}
\tilde{h}_q =  {\cal S}^{-1/2} (h_q - u h_d^{-1} u^\dagger) {\cal S}^{-1/2}, \quad {\cal S}=1+ u h_d^{-2} u^\dagger,
\label{suppeq:eH}
\end{align}
valid when $\varepsilon$ is small, and we have used the rotated basis $\psi = {\cal S}^{1/2} \psi_q$.

Following the above analysis, we now derive the $4 \times 4$ model for the pristine part. From Eq.~(\ref{suppeq:H0}) we have
\begin{align}
h_q  =
\lp
\begin{array}{cccc}
 \epsilon_1 & 0 & v_1^+ & 0 \\
0 & \epsilon_1 & 0 & v_1^- \\
 -v_1^- & 0 & \epsilon_2  & 0 \\
 0 & -v_1^+ & 0 & \epsilon_2 
\end{array}
\rp  ,
\end{align}
\begin{align}
u^\dagger  =
\lp
\begin{array}{cccc}
0 & 0 & v_3^+ & 0 \\
0 & 0 & 0 & v_3^- \\
\end{array}
\rp  ,
\end{align}
and
\begin{align}
h_d = 
\lp
\begin{array}{cc}
\epsilon_3   & 0 \\
0 & \epsilon_3 	 
\end{array}
\rp  ,
\end{align}
where $\epsilon_i = c_{i,0} + c_{i,x} k_x^2 + c_{i,y} k_y^2$, and $v_i^\pm = \pm v_{i, x} k_x + i v_{i, y} k_y  $.
Using $h_d$ and $u$ above, we have
\begin{align}
\tilde{h}_q  =
\lp
\begin{array}{cccc}
\epsilon_c  & 0 & v^+ & 0 \\
0 & \epsilon_c  & 0 & v^- \\
- v^- & 0 & \epsilon_v  & 0 \\
0 & -v^+ & 0 & \epsilon_v
\end{array}
\rp  ,
\label{suppeq:4band}
\end{align}
where $\epsilon_c = \epsilon_1$, and
\begin{align}
\epsilon_v =  
\frac{1}{1+ r} \epsilon_2  - \frac{r}{1+r} \epsilon_3 ,
\quad
v = \sqrt{\frac{1 }{ 1+ r }  } \, v_1  .
\end{align}
Here the $k$-dependent ratio $r$ is
\begin{align}
r = \frac{(v_{3, x} k_x )^2 + (v_{3, y} k_y )^2}{\epsilon_3^2} ,
\end{align}
which controls the renormalization of $\epsilon_2$ and $v_1$. It becomes zero when $v_3 = 0$, i.~e., when $\psi_2$ and $\psi_3$ do not couple with each other, $r = 0$ and there is no renormalization. We note that the form of this pristine part in Eq.~(\ref{suppeq:4band}) is consistent with the model proposed in Ref.~\cite{Qian}; for a full discussion see the section ``Unitary transformation and form of Hamiltonian'' below.

If we re-order the basis to $\{ \psi_{c \uparrow}$, $\psi_{v \uparrow}$; $\psi_{c \downarrow}$, $\psi_{v \downarrow} \}$, this gives the BHZ-type pristine part,
\begin{align}
h_0  = \bar{\epsilon} +
\lp
\begin{array}{cccc}
 m & v^+  & 0 & 0 \\
 - v^- & -m & 0 & 0 \\
 0 & 0 & m  & v^- \\
 0 & 0 & - v^+ & - m 
\end{array}
\rp  ,
\end{align}
where $\bar{\epsilon}= (\epsilon_c + \epsilon_v) / 2$, $m= (\epsilon_c - \epsilon_v) / 2$.
Together with the electric field induced part (neglecting the far away $\psi_3$ band),
\begin{align}
h_1 = 
\lp
\begin{array}{cccc}
\lambda k_y & i \delta_z & \alpha^- & i \delta_x  \\
-i \delta_z & \lambda k_y & i \delta_x & \alpha^- \\
\alpha^+ & -i \delta_x & -\lambda k_y & -i \delta_z  \\
-i \delta_x & \alpha^+ & i \delta_z & -\lambda k_y 
\end{array}
\rp  ,
\end{align}
we obatined the $4 \times 4$ effective hamiltonian near the gap opening
\begin{align}
h^{\rm eff} = h_0 + h_1 .
\end{align}

When focusing on the dispersions and Berry curvatures near the gap opening, we find a convenient estimate for $v$:
\begin{align}
v = \sqrt{\frac{1 }{ 1+ r_0 }  } \, v_1  ,
\quad r_0 = r (\vec k_0) ,
\end{align}
where $\vec k_0 = (0, k_0)$ is the position of the band edge with $k_0 = 0.385~{\rm \AA^{-1}}$. By doing this, we obtained the dispersion and a Berry curvature distribution which agree well with the six band model near the gap opening (see solid and dashed lines in Fig.~\ref{fig1}d and Fig.~\ref{fig1}e for comparison).

\subsection{Berry curvature at the band edge from $4 \times 4$ model\\($\delta_{x,z} = 0$ case)}

Using the $4 \times 4$ band model [see Eq.~(\ref{eq:4band}) and Eq.~(\ref{eq:4bandSOC}) of the main text] to describe 1T'-WTe$_2$, we can express its Berry curvature near the band edge analytically.

Without the electric field induced part $h_1$, the BHZ-type pristine part $h_0$ can be viewed as two decoupled $2 \times 2$ blocks: spin-up block $h_\uparrow$ and spin-down block $h_\downarrow$. The two blocks form a time-reversal pair 
\begin{align}
h_\uparrow (\vec k) = h_\downarrow^* (-\vec k)
=
\bar{\epsilon}_{\vec k} +
\lp
\begin{array}{cc}
 m_{\vec k}  & v_{\vec k}^+  \\
 -v_{\vec k}^- & - m_{\vec k} 
\end{array}
\rp
,
\label{suppeq:GappedDirac}
\end{align}
where $v_{\vec k}^\pm = \pm v_x k_x + i v_y k_y $. 
The two $2 \times 2$ blocks share the same dispersion relation and are spin-degenerate: 
\begin{align}
\varepsilon_{\tau \uparrow\hspace{-0.15em}\downarrow}^{(0)} (\vec k) = \bar{\epsilon}_{\vec k} \pm  \tau \Delta_{\vec k} ,
\quad
\Delta_{\vec k} = \sqrt{(v_x k_x)^2 + (v_y k_y)^2 + m_{\vec k}^2 } ,
\end{align}
with the corresponding eigenstates:
\begin{align}
\psi_{\tau \uparrow }^{(0)}  (\vec k) \sim ( m_{\vec k}  + \tau \Delta_{\vec k}, - v_{\vec k}^-, 0 , 0)^T  , \nn
\psi_{\tau \downarrow }^{(0)} (\vec k) \sim (0,0, m_{\vec k}  + \tau \Delta_{\vec k}, - v_{\vec k}^+ )^T ,
\end{align}
where $\tau = \pm $ denotes conduction or valence band, 
and $2 \Delta_{\vec k}$ is the energy difference between the conduction and valence bands.

When an external perpendicular electric field is applied
and if the spin-orbit couplings induced by the field is much weaker than the atomic spin-orbit couplings ($\alpha_{x,y}, \lambda \ll v_{x,y} $), we then treat $h_1 (\vec k)$ as a small perturbation to the pristine part $h_0 (\vec k)$. We first note that for non-zero $\alpha$, and
within the framework of degenerate perturbation (i.e. neglecting states that are far away in energy), $h_1 (\vec k)$ hybridizes the unperturbed spin up state $\psi_{\tau \uparrow }^{(0)} (\vec k)$ and spin down state $\psi_{\tau \downarrow }^{(0)} (\vec k) $ in the same band $\tau$ into spin split states $\psi_{\tau \xi} (\vec k)$, with higher ($\xi = +1$) or lower ($\xi = -1$) energies:
\begin{align}
\psi_{\tau \xi} (\vec k) = c_{\tau \uparrow}^\xi (\vec k) \psi_{\tau \uparrow }^{(0)}  (\vec k) 
+ c_{\tau \downarrow}^\xi (\vec k) \psi_{\tau \downarrow }^{(0)} (\vec k)  .
\end{align}

Interestingly, in the $\alpha \ll \lambda, \delta_z$ limit, the unperturbed spin up state $\psi_{\tau \uparrow }^{(0)} (\vec k)$ and spin down state $\psi_{\tau \downarrow }^{(0)} (\vec k)$ do not hybridize with each other. Even so, their degeneracies are lifted by 
$\lambda k_y$:
\begin{align}
\psi_{\tau \pm} (\vec k) =  \psi_{\tau \uparrow\hspace{-0.15em}\downarrow}^{(0)}  (\vec k)  , 
\quad 
\varepsilon_{\tau \xi}  (\vec k)
= \bar{\epsilon}_{\vec k}  + \tau \Delta_{\vec k}  + \xi \left| \lambda k_y \right|  .
\end{align}
The Berry curvatures for these non-degenerate states $\psi_{\tau \xi}$ are well defined and read
\begin{align}
\Omega_{\tau \xi}^{(0)} (\vec k) = {\rm sgn}(k_y) \frac{\tau \xi}{2}  
\frac{v_{x} v_{y}}{ \Delta_{\vec k}^3 } (1-\vec k \cdot \nabla ) m_{\vec k}  .
\end{align}

When $\alpha \sim \lambda$, spin up state $\psi_{\tau \uparrow }^{(0)} (\vec k) $ and spin down state $\psi_{\tau \downarrow }^{(0)} (\vec k) $ start to hybridize to form the spin split states $\psi_{\tau \xi}$. The Berry curvature amplitudes for these hybridized states $\psi_{\tau \xi} (\vec k)$ are smaller than that of $\psi_{\tau \uparrow\hspace{-0.15em}\downarrow}^{(0)} (\vec k) $ because the spin up and spin down states have opposite Berry curvatures.

The general form of the Berry curvature $\Omega_{\tau \xi} (\vec k)$ for $\psi_{\tau \xi} (\vec k)$ is complicated. 
Fortunately, along $k_x = 0$ about which the Berry curvature is even, i.~e., $\Omega(k_x,k_y) = \Omega (-k_x, k_y) $, which comes from the TRS and the mirror symmetry in the $y$-direction, its analytical form is greatly simplified:
\begin{align}
\Omega_{\tau \xi} (0, k_y) 
& = |c_{\tau \uparrow}^\xi |^2  \Omega_{\tau \uparrow}^{(0)} (0, k_y) 
+ |c_{\tau \downarrow}^\xi |^2 \Omega_{\tau \downarrow}^{(0)}  (0, k_y)  \nn
& = \lp |c_{\tau \uparrow}^\xi |^2 - |c_{\tau \downarrow}^\xi |^2 \rp \Omega_{\tau \uparrow}^{(0)} (0, k_y) .
\end{align}
where $c_{\tau \uparrow}^\xi$ and $c_{\tau \downarrow}^\xi$ satisfy
\begin{align}
|c_{\tau \uparrow}^\xi |^2 - |c_{\tau \downarrow}^\xi |^2  = \tau \xi \,  \frac{\lambda \, {\rm sgn}(k_y)}{(\alpha_y^2 + \lambda^2)^{1/2}}.
\end{align}
With this, the Berry curvature $\Omega_{\tau \xi} (\vec k)$ at the band edge $\vec k' = (k_x = 0, k_y )$ is
\begin{align}
\Omega_{\tau \pm } (\vec k') & =  \frac{\lambda {\rm sgn}(k_y)}{ (\lambda^2 + \alpha_y^2)^{1/2} } \frac{\tau \xi}{2} \frac{v_x v_y}{ \Delta_{\vec k'}^3 } (1- \vec k' \cdot \nabla ) m_{\vec k'} \nn
& =  \frac{\lambda}{ (\lambda^2 + \alpha_y^2)^{1/2} } \Omega_{\tau \uparrow\hspace{-0.15em}\downarrow }^{(0)} (\vec k').
\end{align}

\subsection{Berry curvature and magnetic moment distribution\\($\lambda = \alpha_{x,y} = \delta_x = 0$,
$\delta_z \neq 0$ case)}

\begin{figure}[t!]
\includegraphics[width=1\columnwidth]{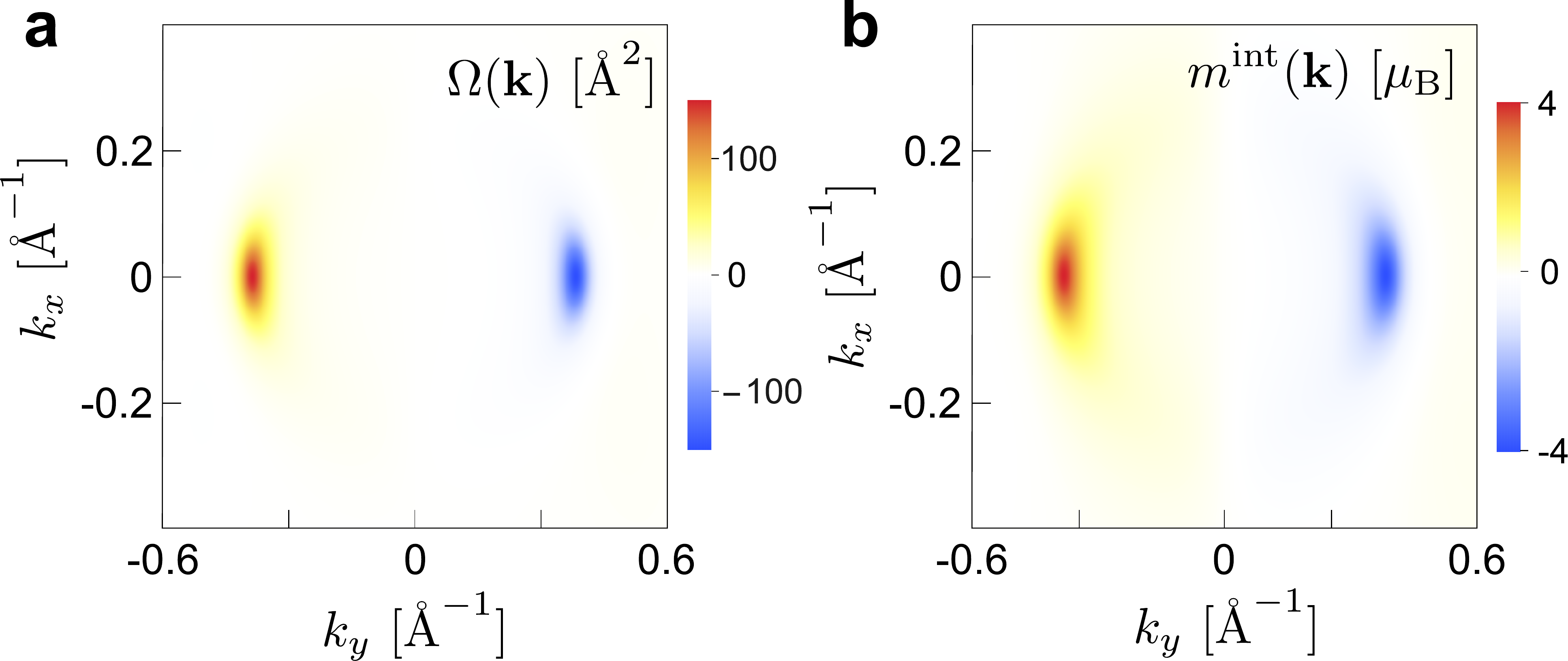}
\caption{{\bf a},{\bf b} Berry curvature and magnetic moment distribution for the spin split conduction band with lower energy in $k$-space. Parameters for the pristine part $H_0$ are listed in Table~\ref{tab:parameters}, while for $H_1$ we used $\alpha, \lambda, \delta_x=0$,  and $\delta_z = 0.025~{\rm eV}$.}
\label{fig-s2}
\end{figure}

Here we show the anisotropic Berry curvature and magnetic moment distribution in $k$-space away from the gap opening, using the six band description (SBD) we developed (see Fig.~\ref{fig-s2}). 
The formula for calculating Berry curvature and moment distribution are
\begin{align}
\Omega_n (\vec k)= 
2 \Re \sum_{n' \neq n}
\frac{i \la n | \p H / \p k_x  | n' \ra \la n' | \p H / \p k_y  | n \ra }
{ ( \varepsilon_n - \varepsilon_{n'} )^2 }  ,
\label{eq:bc}
\end{align}
\begin{align}
m_{n}^{\rm int} (\vec k) = 
\frac{e}{\hbar} \Re \sum_{n' \neq n}
\frac{i \la n | \p H / \p k_x  | n' \ra \la n' | \p H / \p k_y  | n \ra }
{ \varepsilon_n - \varepsilon_{n'} },
\label{eq:omm}
\end{align}
where $n = \{\tau \xi\}$ is the short-hand form, and $H$ is the full six band hamiltonian.

\subsection{Non-linear anomalous Hall effect}

Just as $m_n^{\rm int}(\vec k)$ 
discussed in the main text gives rise to ME, $\Omega_n (\vec k)$ in the bands enable 1T'-WTe$_2$ to exhibit a quantum non-linear Hall effect at zero magnetic field. 
This can be seen under general symmetry considerations~\cite{Sodemann}. For the convenience of the reader, we outline this symmetry analysis for a 2D system that only has in-plane mirror symmetry (e.g., 1T'-WTe$_2$). The non-linear Hall current can be written as  
$j_a = \chi_{abb} E_b E_b$ ($a,b= x,y$). Under the
operation $(x,y) \to (x, -y)$, we have $(j_x, j_y) \to (j_x, -j_y)$ and $(E_x, E_y) \to (E_x, -E_y)$. This allows a non-zero $\chi_{x y y}$ (while $\chi_{yxx}$ vanishes). Under an in-plane DC electric field, $\chi_{x y y}$ can be obtained using~\cite{Sodemann}
\begin{align}
\chi_{x y y} &=    \frac{\tau}{2}
\frac{e^3}{\hbar^2} \sum_{n, \vec k} \, f_{n \vec k}^{(0)} \,  \p_{k_y} \Omega_n (\vec k), 
\label{eq:chi}
\end{align}
where $\tau$ is the transport scattering time. 
Similar to Eq.~(\ref{eq:MzJ}) above, $\Omega_n (\vec k)$ is an odd function of $k_y$ [see Fig.~\ref{fig-s2}(a)]. As a result, nonzero $\chi_{a b b}$ only occurs when $a = x$ and $b = y$: only electric field along $y$ induces a non-linear Hall effect along $x$. 
When $\vec E$ is parallel to the $x$-direction, the nonlinear Hall effect (as well as the kinetic ME effect) vanishes. 

\begin{figure}[t!]
\includegraphics[scale=0.2]{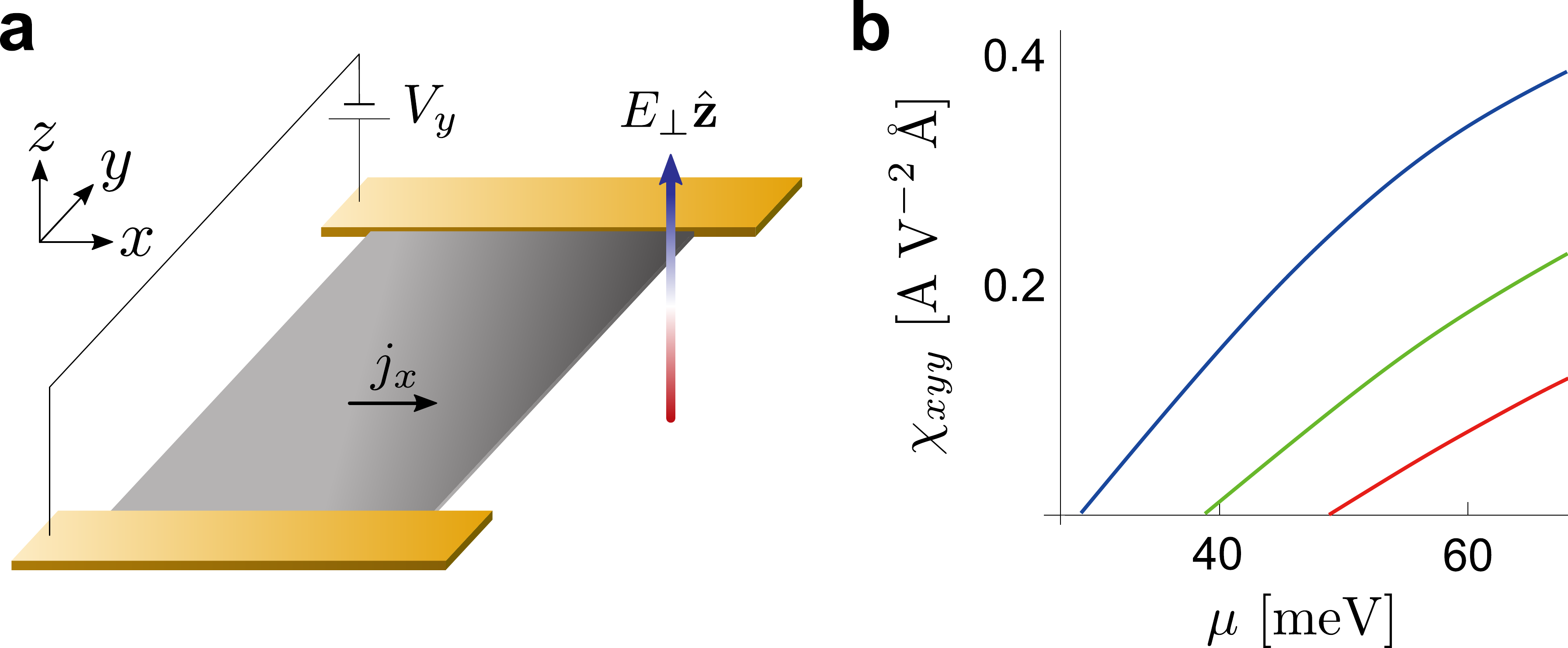}
\caption{{\bf a} Schematic of a 1T'-WTe$_2$ monolayer under a perpendicular electric field $E_\perp \hat{\vec z}$ and an in-plane electric field $E_y$, which can give rise to a non-linear Hall current $j_x$. 
({\bf b}) Calculated non-linear Hall conductivity $\chi_{xyy}$ (blue) from Eq.~(\ref{eq:chi}) using $\tau = 50~{\rm fs}$. Parameters used for the pristine part are the same as those in Fig.~\ref{fig1}; for the electric field induced part we used
$\alpha,\lambda,\delta_x = 0$, and $\delta_z = 0.075~{\rm eV}$ (blue), $0.05~{\rm eV}$ (green), and $0.025~{\rm eV}$ (red).}
\label{fig-s3}
\end{figure}

To illustrate the quantum non-linear Hall effect, we numerically integrate Eq.~(\ref{eq:chi}) to obtain a finite non-linear
Hall current conductivity $\chi_{xyy}$ in Fig.~\ref{fig-s3}b 
using a scattering time $\tau = 50~{\rm fs}$. Similar to $M_z$ in the main text,
we used the SBD description in order to capture the full reciprocal space distribution of the Berry curvature. This non-linear Hall conductivity can be probed in a conventional Hall bar measurement (Fig.~\ref{fig-s3}) and provides a fully electrical way of mapping the Berry curvature (dipole). 

\subsection{Unitary transformation and form of Hamiltonian}

We note that the model in the supplement of Ref.~\cite{Qian} is equivalent to our model for the pristine part. For the convenience of the reader, we reproduce 
the four band hamiltonian in Ref.~\cite{Qian} as 
\begin{align}
H_F = 
\lp
\begin{array}{cccc}
\epsilon_c & 0 & - i v_x k_x &  v_y k_y  \\
0 & \epsilon_c & v_y k_y & -i v_x k_x \\
i v_x k_x & v_y k_y & \epsilon_v & 0  \\
v_y k_y & i v_x k_x & 0 & \epsilon_v 
\end{array}
\rp .
\end{align}
To see the equivalence, we use the unitary transformation
\begin{align}
U =  \frac{1}{\sqrt{2}}
\lp
\begin{array}{cccc}
1 & 1 & 0 &  0  \\
1 & -1 & 0 & 0 \\
0 & 0 & i & i  \\
0 & 0 & i & -i 
\end{array}
\rp.
\end{align}
Applying the unitary transformation, we have
\begin{align}
U^\dagger H_F U =  \tilde{h}_q ,
\end{align}
reproducing Eq.~(\ref{suppeq:4band}).

\clearpage

\end{document}